\documentclass[11pt,a4paper]{article}

\usepackage{a4wide}
\begin{document}

\begin{titlepage}
\begin{flushright}
hep-th/9610241
\\
ZU-TH-18/96
\\
FSU-TPI-11/96
\end{flushright}
\vspace{0.4cm}

\begin{center}
\huge{\bf Finite Temperature Schwinger Model\\
with\\
Chirality Breaking Boundary Conditions}
\end{center}

\vspace{0.6cm}
\begin{center}
{\bf S. D\"urr}\\
\vspace{0.2cm}
{\sl Institute for Theoretical Physics\\
University of Z\"urich (Irchel)\\
CH - 8057 Z\"urich (Switzerland)}\\
{\small\tt email: duerr@physik.unizh.ch}\\
\vspace{0.8cm}
{\bf A. Wipf}\\
\vspace{0.2cm}
{\sl Theoretisch-Physikalisches Insititut\\
Friedrich-Schiller University of Jena\\
D - 07743 Jena (Germany)}\\
{\small\tt email: aww@hpfs1.physik.uni-jena.de}
\end{center}

\vspace{0.8cm}

\begin{abstract}\noindent
The $N_f$-flavour Schwinger Model on a finite space $0\leq x^1\leq L$ and
subject to bag-type boundary-conditions at $x^1=0$ und $x^1=L$ is solved
at finite temperature $T=1/\beta$.
The boundary conditions  depend on a real parameter $\theta$ and break
the axial flavour symmetry.
\newline
We argue that this approach is more appropriate to study the broken phases
than introducing small quark masses, since all calculations can be performed
analytically. 
\newline
In the imaginary time formalism we determine the thermal correlators for
the fermion-fields and the determinant of the Dirac-operator in arbitrary
background gauge-fields. 
We show that the boundary conditions induce a $C\!P$-odd $\theta$-term
in the effective action.
\newline
The chiral condensate, and in particular its $T$- and $L$- dependence, is
calculated for $N_f$ fermions.
It is seen to depend on the order in which the two lengths $\beta=1/T$
and $L$ are sent to infinity.
\end{abstract}

\vspace{0.6cm}
\end{titlepage}

\newcommand{\pas}{\partial\!\!\!/}
\newcommand{\psl}{\partial\!\!\!/}
\newcommand{\Dsl}{D\!\!\!\!/}
\newcommand{\lrar}{\longrightarrow}
\newcommand{\pa}{\partial}
\newcommand{\psb}{\overline{\psi}}
\newcommand{\psd}{\psi^{\dagger}}
\newcommand{\etd}{\eta^{\dagger}}
\newcommand{\chd}{\chi^{\dagger}}
\newcommand{\tr}{\,{\rm tr}\,}
\newcommand{\Tr}{\,{\rm Tr}\,}
\newcommand{\til}{\tilde}
\renewcommand{\dag}{^\dagger}
\newcommand{\pri}{^\prime}
\newcommand{\hb}{\hbar}
\renewcommand{\>}{\rangle}
\newcommand{\ran}{\rangle}
\newcommand{\<}{\langle}
\newcommand{\lan}{\langle}
\newcommand{\gaf}{\gamma_5}
\newcommand{\lap}{\triangle}
\newcommand{\paw}{\par} 
\newcommand{\pan}{\newline}
\newcommand{\uad}{\ }
\newcommand{\al}{\alpha}
\newcommand{\be}{\beta}
\newcommand{\ga}{\gamma}
\newcommand{\de}{\delta}
\newcommand{\ep}{\epsilon}
\newcommand{\ve}{\varepsilon}
\newcommand{\ze}{\zeta}
\newcommand{\et}{\eta}
\renewcommand{\th}{\theta}
\newcommand{\vt}{\vartheta}
\newcommand{\io}{\iota}
\newcommand{\ka}{\kappa}
\newcommand{\la}{\lambda}
\newcommand{\rh}{\rho}
\newcommand{\vr}{\varrho}
\newcommand{\si}{\sigma}
\newcommand{\ta}{\tau}
\newcommand{\ph}{\phi}
\newcommand{\vp}{\varphi}
\newcommand{\ch}{\chi}
\newcommand{\ps}{\psi}
\newcommand{\om}{\omega}
\newcommand{\ov}{\over} 
\newcommand{\cd}{{\cal D}}
\newcommand{\rch}{{\rm ch}}
\newcommand{\rsh}{{\rm sh}}
\newcommand{\msh}{\mbox{sh}}
\newcommand{\mch}{\mbox{ch}}
\newcommand{\mssh}{\mbox{\small sh}}
\newcommand{\msch}{\mbox{\small ch}}
\newcommand{\beq}{\begin{equation}}
\newcommand{\eeq}{\end{equation}}
\newcommand{\bdm}{\begin{displaymath}}
\newcommand{\edm}{\end{displaymath}}
\newcommand{\bea}{\begin{eqnarray}}
\newcommand{\eea}{\end{eqnarray}}
\newcommand{\bes}{\begin{eqnarray*}}
\newcommand{\ees}{\end{eqnarray*}}

\def\ha{{1\over 2}}
\def\pr{\prime}

\normalsize


\section{Introduction}


Over the past decades the Schwinger model \cite{O.0} has proved to be an
excellent laboratory for field theory because it turned out to shed some light
on a couple of questions which naturally arise in realistic gauge-field
theories, but lead to immense difficulties as soon as one tries to attack
them directly. Longstanding problems of this type are the wellknown
$U(1)_A$-problem \cite{HeHoItAdFr},
the question whether QCD in the chiral limit 
shows a spontaneous breakdown of the chiral symmetry and the question about
the nature of the chiral phase transition at $\sim 200$ MeV 
\cite{ShBook,GrSeSoTi}. 

The Schwinger model is known to be the most simple model field theory which
exhibits chiral symmetry breaking. However the quantization on the plane
suffers from the deficit, that a naive calculation of 
the condensates $\<\psb\psi\>$ and 
$\<\psb\gaf\psi\>$ gives zero results and the correct values can be derived
only a posteriori by using the clustering theorem \cite{GattringerSeiler}.
Whenever a symmetry is expected to be broken it is most recommendable to break
it explicitly and to try to determine how the system behaves in the limit when
the external trigger is softly removed.
Thus it is most natural for both the Schwinger model and QCD to break
explicitly the axial flavour symmetry and to investigate how observables do
behave in the limit where the symmetry is restored.

The most direct way to do this is to introduce small fermion masses
and to try to determine how the chiral correlators behave in the limit where
these fermion masses tend to be negligible as compared to the intrinsic energy
scale of the gauge-interaction.
Once these calculations do predict nonvanishing chiral
condensates in the thermodynamic
limit one can be sure that a spontaneous breaking of the axial flavour
symmetry $SU(N_f)_A$ really takes place - however this is not a conditio sine
qua non.
There is however a technical obstacle to this approach:
the value of the chiral condensates is related to the mean 
level density of the eigenvalues of the Dirac operator 
\cite{BanksCasher} in the infrared. 
Unfortunately, the spectral density of the massive Dirac operator is
only known for very special background gauge-fields. 

In this paper we shall break the chiral symmetry 
explicitly by boundary conditions for the
fermions instead of giving them a small mass. Although this version seems at
first sight less natural, it has many advantages - both conceptual and
calculational in nature. The most important point is clearly the fact that it
allows for an entirely analytical treatment.
In a previous paper \cite{WiDu} we investigated QCD-type theories
with $N_f$ massless flavours on an even-dimensional ($d\!=\!2n$) euclidean
manifold $M$ with boundary $\pa M$ on which the boundary conditions
studied by Hrasko and Balog \cite{HrBa} have been applied.
These chirality-breaking- (CB-) boundary-conditions relate the different
spin components of one flavour on $\pa M$ and are neutral with respect to
vector-flavour transformations - so that the (gauge-invariant) fermionic
determinant is the same for all flavours.
For a simply connected $M$, e.g. a ball, the instanton
number, which in four dimensions takes the form
\beq
q={1\ov32\pi^2}\int F_{\mu\nu}^a\;\til F_{\mu\nu}^a\;d^4x,
\label{intr.1}
\eeq
is {\em not quantized} and may take any real value \cite{WiDu}. 
Contrary to the situation on a compact manifold without boundary, on 
which $q$ is integer \cite{O.4}, the configuration space is
{\em topologically trivial} (i.e. without disconnected instanton sectors)
\cite{WiDu}.
In addition there are {\em no fermionic zero modes} \cite{WiDu,O.5} which
usually tend to complicate the quantization considerably \cite{SaWi}.

Our previously cited work focused on the euclidean 
$N_f$-flavour $U(N_c)$- or $SU(N_c)$- gauge-theories inside $2n$-dimensional
balls of radius $R$. We computed that part of the
effective action reflecting the interaction of the particles with the
boundary $S^{2n-1}_R$.
Here we investigate whether the approach of breaking the
$SU(N_f)_A$ symmetry by  boundary conditions can
be extended to gauge-systems in thermal equilibrium
states. In the imaginary time formalism 
spacetime is then a cylindrical manifolds 
$M=[0,\beta]\times \{\mbox{space}\}$ and (anti)periodic
boundary conditions for the (fermi)bose-fields in
the euclidean time $x^0$ with period $\beta=1/T$
are imposed.

Note that at finite temperature it is only the
boundary of space and not of space-time  where chirality is
broken and it is a priori an open question whether this is sufficient to
trigger a chiral symmetry breaking  even in the one flavour case.
In addition there is a technical obstacle to extending the CB-boundary-condition
approach to non-simply connected manifolds, e.g a cylinder. On cylinders
the standard decomposition for the Dirac operator, on which the
analytic treatment heavily relies, must be modified.
The present paper is a technical one - mostly devoted to show how this
difficulty can be overcome. From the physical point of view it is our aim to
investigate how the breakdown of the chiral symmetry - when triggered by
boundary conditions - in the one- and the multi-flavour cases is affected by
finite temperature effects.

Here we shall quantize the Schwinger Model with action
\beq
\begin{array}{c}
S[A,\psd,\ps]=S_B[A]+S_F[A,\psd,\ps]
\\
\\
S_B={1\ov 4}\int\limits_M  F_{\mu\nu} F_{\mu\nu}\quad,\quad
S_F=\sum\limits_{n=1}^{N_f} \int\limits_{M}\psd_n i\Dsl\;\ps_n
\end{array}
\label{mfsm.1}
\eeq
on the manifold
\beq
M=[0,\be]\times[0,L]\quad\ni\quad(x^0,x^1)
\label{pro.1}
\eeq
with volume $V=\be L$. At finite temperature the fields $A$ and $\psi$ are
periodic and antiperiodic in euclidean time with period $\beta$
and hence $x^0=0$ and $x^0=\beta$ are identified. This means that
$[0,\beta]\times[0,L]$ is a cylinder with circumference $\beta=
1/T$.
At the spatial ends of the cylinder (i.e. at $x^1=0$ and
$x^1=L$) specific CB-boundary-conditions are applied.
Then there are no fermionic
zero modes (see next section) and the generating functional for the
fermions in a given gauge-field background $A$ is given by the textbook formula
\beq
Z_F [A,\etd,\et]\;=\;
\det(i\Dsl)\;\;e^{\,\int\etd(i\Dsl)^{-1}\et}.
\label{mfsm.3}
\eeq

We shall see that these CB-boundary-conditions indeed generate chiral
condensates for any finite length $L$
of the cylinder.
However in the limit $\be\rightarrow\infty,L\rightarrow\infty$
the condensates will only survive for the one flavour case and this only if the
limit $\be\rightarrow\infty$ is taken before the limit $L\rightarrow
\infty$.

During the calculations the following abbreviations are
used for notational simplicity:
\beq
\tau={\be\ov 2L}\quad,\qquad 
\eta={x^1+y^1\ov L}\quad,\qquad \xi={x^1\ov L}.
\label{abbrev}
\eeq

This paper is organized as follows :
In section 2 we discuss the
CB-boundary-conditions to be applied together with 
some immediate consequences for
the spectrum of the Dirac operator $i\Dsl$.
Section 3 is devoted to the question of how to decompose an
arbitrary gauge-field on a cylinder.
In section 4 we compute the fermionic Green's function 
with respect to CB-boundary-conditions in arbitrary external fields.
In section 5 we determine the effective action after the fermions have been
integrated out.
Using the results of the two previous steps the chiral condensates are
calculated in section 6.
In section 7 we show that the value of the chiral condensate crucially
depends not only on the number of flavours but also on the order in which the
two limits $\be\rightarrow\infty$ and $L\rightarrow\infty$ are
performed.
Finally we compare our result with the condensate generated by
fractons on a torus of identical size and with analogous results of
noncommutativity of the limits $m\rightarrow 0$, $L\rightarrow\infty$ in
the the usual small-quark-mass approach.
In the appendices we derive the boundary Seeley-DeWitt coefficient used in
the body of the paper.


\section{Chirality Breaking Boundary Conditions}


In this section we shall shortly review the boundary conditions as discussed
by Hrasko and Balog \cite{HrBa} together with their most important consequences
\cite{WiDu,FaGaMuSaSo}.

Since $Z_F$ should be real we want $i\Dsl\;$ to be symmetric under the scalar
product
\bdm
(\ch,\ps) := \int\limits_M \, \chd\ps
\edm
from which we get the condition
\beq
(\ch,i\Dsl\,\ps)-(i\Dsl\,\ch,\ps)=i\oint\limits_{\pa M}\chd\ga_n\ps\equiv 0\ .
\label{hbbc.1}
\eeq
Imposing local linear boundary conditions which ensure this requirement
amounts to have $\chd\ga_n\ps=0\ $ on $\pa M$ for each pair,
which is achieved by
\beq
\ps=B\ps\quad\mbox{on}\quad\pa M\qquad\qquad\mbox{with}\qquad\qquad
B\dag\ga_n B=-\ga_n\quad,\quad B^2=1,
\label{hbbc.2}
\eeq
where $\ga_n=(\ga,n)=n_\mu \ga_\mu=n\!\!\!\slash\;$ and $n_\mu$
is the outward oriented normal vectorfield on $\pa M$.
We shall choose the one-parametric family of boundary operators \cite{HrBa}
\beq
B\equiv B_\th:\equiv i\gaf e^{\th\gaf}\ga_n
\label{hbbc.3}
\eeq
which is understood to act as the identity in flavour space.
These CB-boundary-conditions break the $\gaf$ invariance of the theory,
making the $N_f$ flavour theory invariant under $SU(N_f)_V$ instead of 
$SU(N_f)_L\,\times SU(N_f)_R$. 
Later they will be supplemented by suitable boundary conditions for the
gauge-field. 
These boundary conditions imply that there is no net 
$U(1)$-current leaking through the boundary, 
since $n\cdot j=\psd\ga_n\ps=0$ on $\pa M$.

In the following we shall make use of a Feynman Hellmann \cite{O.20}
boundary formula, which may be derived from (\ref{hbbc.1},\ref{hbbc.2},\ref{hbbc.3}) 
\cite{WiDu}
\beq
{d\ov d\th}\la_k={i\ov2}\oint\ps\dag_k(\ga\cdot n)\gaf\ps_k=
-\la_k(\ps_k,\gaf\ps_k)\ ,
\label{hbbc.4}
\eeq
where the $\la_k$ denote the eigenvalues of $i\Dsl$.

We choose the chiral representation
$\ga_0=\si_1, \ga_1=\si_2$ and 
$\gaf=\si_3$
. 
Then the boundary operators at the two ends of the cylinder read
\beq
B_L=-\left(\begin{array}{cc}0&e^{\th}\\
e^{-\th}&0\end{array}\right)\;({\mbox{at $x^1\!=\!0$}})
\quad\hbox{and}\quad
B_R=+\left(\begin{array}{cc}0&e^{\th}\\
e^{-\th}&0\end{array}\right)\;({\mbox{at $x^1\!=\!L$}})\ .
\label{hbbc.6}
\eeq
The most important properties of these boundary conditions are summarized as
follows \cite{WiDu}:
\newline (1)
The Dirac operator has a {\em purely discrete} real spectrum which is
{\em not symmetric} with respect to zero.
\newline (2)
The Dirac operator has {\em no zero modes}.
\newline (3)
The instanton number $q=\frac{1}{4\pi}\int\ep_{\mu\nu}F_{\mu\nu}=
\frac{1}{2\pi}\int E$ is {\em not quantized\ }.
The second property allows us to 
calculate expectation values of gauge-invariant operators as
\bea
\<O\>&=&\int{\<O\>}_A\ d\mu_\th[A],\quad\hbox{where}
\label{mfsm.5}
\\
d\mu_\th[A]&=&{1\ov Z_F}\;{\det}_\th(i\Dsl\,)\;e^{-S_B[A]}\ D[A].
\label{mfsm.6}
\eea
Here $D[A]$ is assumed to contain the gauge-fixing factor including the
corresponding Fadeev-Popov determinant and $\<O\>_A$ denotes the expectation
value of $O$ in a fixed background
\beq
\<O\>_A={1\ov\det_\th(i\Dsl)}\cdot\int\cd\psd\cd\ps\ \;O\
e^{-\int\psd i\Dsl\;\ps}.
\label{mfsm.7}
\eeq
Throughout $\th$ is the free parameter in boundary
operators (\ref{hbbc.6}). We shall see that
the $\th$-dependence of 
the fermionic determinant $\det_\th(i\Dsl)$ can be calculated analytically.


\section{Decomposition and Deformation techniques}


In this section we present the decomposition and deformation techniques
needed to determine the functional determinant of the Dirac operator on the
cylinder with CB-boundary-conditions as given by (\ref{hbbc.2}) and
(\ref{hbbc.6}) .

On simply connected regions we have the
decomposition $eA_\mu=-\ep_{\mu\nu}\pa_\nu\ph+
\pa_\mu\chi$ such that $eF_{01}=\triangle \phi$.
On the cylinder there is a one to one  correspondence 
between $\phi$ and $eF_{01}$ if $\phi$ obeys
Dirichlet boundary conditions at the two ends of the
cylinder. But cylinders are not simply connected, $\pi_1(M)={\bf Z}$,
and as a result the Polyakov-loop operators
\bdm
e^{ie\int_0^\be A_0\;dx^0}=:e^{2\pi ic},
\edm
i.e. $e\int_0^\beta A_0$ mod $2\pi$, are gauge-invariant. 
On the other hand, using the $\beta$-periodicity of $\chi$ and
the Dirichlet boundary conditions on $\phi$, the above
decomposition would imply that $\int_0^\be\int_0^L A_0=0$,
a condition which does not hold in general (take a constant $A_0$). This
simple observation already indicates, that 
the correct decomposition of $A_\mu$ on the cylinder reads
\beq
\begin{array}{rrr}
eA_0=&-\pa_1\ph+\pa_0\ch&+\frac{2 \pi}{\be}c
\\
eA_1=&+\pa_0\ph+\pa_1\ch ,&{}
\end{array}
\label{dadt.11}
\eeq
where $\ph$ obeys Dirichlet boundary conditions at $x^1=0,L$ and $\ch$
fulfills $\ch(0)+\ch(L)=0$ and $c\in[0,1[$ is the constant harmonic part.
To prove (\ref{dadt.11}) one Fourier decomposes the various
fields and carefully handles the zero-modes of the
Laplacian. The harmonic part can then be reconstructed from
its values on the boundaries.\pan
The Dirac operator $i\Dsl=i\ga_\mu(\pa_\mu-ieA_\mu)$ may be factorized
according to
\beq
i\Dsl=G\dag i\Dsl_{\,0} G,\quad\hbox{where}\quad
i\Dsl_{\,0}=\gamma^0(i\pa_0+2\pi c/\beta)+\gamma^1i\pa_1,\label{dadt.15}
\eeq
and
\beq
G=\left(\begin{array}{cc}
g^{*-1}&0\\0&g\end{array}\right),\quad
g:\equiv e^{-(\ph+i\ch)}.
\label{dadt.16}
\eeq
The prepotential $g$ is an element of the complexified gauge-group
$U(1)^{*}=S^1\times R_+$.
\noindent
Now we deform the prepotential and Dirac operator as
\beq
g_\al:\equiv e^{-\al(\ph+i\chi)}\quad\hbox{and}\quad
i\Dsl_\al=G\dag_\al\,i\Dsl_{\,0}\;G_\al
\label{dadt.18}
\eeq
such the deformed operator interpolates between the
free and full ones:
$i\Dsl_{\al=1}=i\Dsl$ and $i\Dsl_{\al=0}=i\Dsl_{\,0}$.
By using
\beq
{d\ov d\al}G_\al=-G_\al H \quad ,\quad H=
\left(\begin{array}{cc}-h^{*}&0\\0&h\end{array}\right)=
-\ph\gaf+i\ch I\ .
\label{dadt.20}
\eeq
one finds for the $\al$-variation of the integrated
heatkernel of $(i\Dsl_\al)^2$
\beq
{d\ov d\al}(\tr\{e^{t\Dsl_\al^2}\})
=2t\tr\{e^{t\Dsl_\al^2}(H+H\dag)(i\Dsl_\al)^2\}
=2t{d\ov dt}(\tr\{e^{t(\Dsl_\al)^2} 2\ph\gaf\})
\label{dadt.21}
\eeq
and this formula will prove to be useful in section 5.


\section{Fermionic Propagator w.r.t. Boundary Conditions}


In order to calculate the condensates we need
the Green's function $S_\th$ of the Dirac operator
$i\Dsl$ on the cylinder subject to the CB-boundary conditions. 
This Green's function obeys
\bea
(i\Dsl\;S_\th) (x,y)&=&\de(x-y)
\label{fpro.1}\\
S_\th(x^0\!+\!\be,x^1,y)&=&-\,S_\th(x,y)
\label{fpro.2}\\
(B_L \;S_\th)(x^0,x^1\!=\!0,y)&=&
S_\th(x^0,x^1\!=\!0,y)
\label{fpro.3}\\
(B_R \;S_\th)(x^0,x^1\!=\!L,y)&=&
S_\th(x^0,x^1\!=\!L,y)
\label{fpro.4}
\eea
plus the adjoint relations with respect to $y$. The dependence of the
gauge-potential has not been made explicit and the boundary operator
$B_{L/R}$ is the one defined in (\ref{hbbc.6}).
From the factorization property (\ref{dadt.15}) for the Dirac operator it
follows at once, that $S_\th$ is related to the
Green's function $\til S_\th$ of $i\Dsl_{\,0}$ as
\beq
S_\th(x,y)=G^{-1}(x) \til S_\th(x,y) G^{\dagger\;-1}(y)\ .
\label{fpro.5}
\eeq
Indeed, since the field $\ph$ obeys Dirichlet boundary conditions 
at the ends of the cylinder, $g$ is unitary there and the boundary conditions
(\ref{fpro.2}-\ref{fpro.4}) transform into the identical ones
for 
\bdm
\til S_\th(x,y)=
\left(\begin{array}{cc} \til S_{++}& \til S_{+-}\\
\til S_{-+}&\til S_{--}\end{array}\right),
\edm
where the indices refer to chirality.
\pan
The free Green's function on the cylinder of infinite length
\bdm
\til S_{ther}(x,y)=
{1\ov 2\pi i}\sum_{n\in Z} (-1)^n 
e^{2\pi ic(\xi^0\!-\!n\be)/\be}\cdot
\pmatrix{\displaystyle 0&\displaystyle{1\ov \xi^0+i\xi^1-n\beta}\cr 
\displaystyle {1\ov\xi^0-i\xi^1-n\beta}&\displaystyle 0},
\edm
where $\xi^\mu=x^\mu-y^\mu$, 
is purely off-diagonal and thus chirality preserving, 
as expected, since thermal boundary conditions are
chirality-neutral.
To implement the chirality-breaking boundary conditions
at the ends of the cylinder on can 
either augment $\til S_{ther}$ by pieces built from the zero
modes (which themselves cannot obey the L/R conditions simultaneously)
or by exploiting analyticity arguments.
In either case we end up with
\bea
\til S_\th(x,y)={i\ov 2\pi}\cdot
\sum_{m,n\in Z\times Z}(-1)^{(m+n)}\cdot
e^{2\pi ic(\xi^0/\be\!-\!n)}
\cdot
\pmatrix{e^\theta/r_{nm}&-(1/s_{nm})\cr
         -(1/\bar s_{nm})&e^{-\theta}/\bar r_{nm}},
\label{fpro.6}
\eea
where $r_{nm}=\xi^0+i\eta-(n\beta+2imL)$ and $s_{nm}$ is the
same expression with $\eta\equiv x^1+y^1$ replaced by $\xi^1$.
From this explicit expression one sees at once that
the off-diagonal elements only depend on $x^\mu\!-\!y^\mu$ and become singular
for $x\rightarrow y$ , whereas the diagonal elements depend on both
$x^\mu$ and $y^\mu$ separately but are regular at coinciding points inside the
cylinder.
The sum over $m$ respectively $n$ in (\ref{fpro.6}) can be performed 
by using \cite{GrRy}
\beq
\sum\limits_Z (-1)^m{e^{imx}\ov m+i a}=-{i\pi\ov \sinh a\pi}e^{ax}
\qquad\qquad (-\pi\leq x\leq \pi)
\label{serie}
\eeq
with the results
\bea
\til S_\th(x,y)={ie^{2\pi i c\xi^0/\beta}\ov 4L}\cdot\sum_Z(-1)^n
e^{-2\pi i nc}\cdot
\pmatrix{\displaystyle {e^{\theta}\ov\sinh(\pi r_{n0}/2L)}&
\displaystyle -{1\ov\sinh(\pi s_{n0}/2L)}\cr
\displaystyle -{1\ov\sinh(\pi\bar s_{n0}/2L)}&
\displaystyle {e^{-\theta}\ov\sinh(\pi\bar r_{n0}/2L)}},
\label{fpro.7a}
\eea
or
\bea
\til S_\th(x,y)={ie^{2\pi ic \xi^0/\be}\ov 2\beta}\cdot\sum_Z(-1)^m
\cdot
\pmatrix{\displaystyle{e^{\theta}
e^{-2\pi i c r_{0m}/\be}\ov\sin(\pi r_{0m}/\be)}&
\displaystyle -{e^{-2\pi i c s_{0m}/\be}\ov\sin(\pi s_{0m}/\be)}\cr
\displaystyle -{e^{-2\pi i c \bar s_{om}/\be}\ov \sin(\pi\bar s_{0m}/\be)}&
\displaystyle {e^{-\theta}
e^{-2\pi i c \bar r_{om}/\be}\ov\sin(\pi\bar r_{0m}/\be)}}\ .
\label{fpro.7b}
\eea
valid for $c\in[-\ha,\ha]$.
For calculating the chiral condensates we shall need 
the $++$ and $--$ elements at 
coinciding points inside the cylinder.
From (\ref{fpro.7a}) we find the expression
\beq
\til S_\th(x,x)_{\pm\!\pm}=
\pm{e^{\pm\th}\ov 4L}\sum_{n\in Z}(-)^n
 {e^{\pm 2in\pi c}\ov \sin(\pi[\xi-in\tau])}.
\label{fpro.8}
\eeq
which rapidly converges for low temperature, and from
(\ref{fpro.7b}) the alternative form
\beq
\til S_\th(x,x)_{\pm\!\pm}=\pm
{e^{\pm\th}\ov2\be}\sum_{m\in Z}(-)^m
{e^{\pm 2\pi c(\xi+m)/\tau}
\ov\sinh(\pi[\xi+m]/\tau)}
\label{fpro.10}
\eeq
which is adequate for high temperature.

With (\ref{fpro.5}) we end up with the following 
expressions for chirality violating entries of
the fermionic Green's function on the diagonal
\beq
S_\th(x;x)_{\pm\!\pm}=
e^{\mp2\ph(x)}\til S_\th(x;x)_{\pm\!\pm}\qquad.\label{fpro.11}
\eeq
The free Green's functions $\tilde S_{\pm\pm}$ have been computed
in (\ref{fpro.8}) and (\ref{fpro.10}). They depend only
on the harmonic part $c$ in the decomposition (\ref{dadt.11})
of the gauge-potential.


\section{Fermionic Determinant w.r.t. Boundary Conditions}


In this section we shall compute the $\th$-dependence of the fermionic
determinant. We shall see that the scattering of the fermions off the
boundary generates a CP-odd $\th$-term in the effective action
for the gauge-bosons.


\subsection{Zetafunction Definition}

The Dirac operator and the boundary conditions are both flavour neutral.
Thus the determinant is the same for all flavours and it is sufficient to
calculate it for one flavour. For the explicit calculations we shall use
the gauge-invariant $\ze$-function definition of the determinant \cite{O.21,
ElOdBook}
\beq
\log\det{}_\th(i\Dsl):\equiv{1\ov2}\log\det{}_\th(-\Dsl^2):\equiv
-{1\ov2}{d\ov ds}\bigg\vert_{s=0}\ze{}_\th(-\Dsl^2,s)
\label{fdet.1}
\eeq
and calculate the $\th$-dependence of the $\ze$-function by means of the
boundary Feynman Hellmann formula (\ref{hbbc.4}). Denoting $\{\mu_k\vert
k\in {\bf N}\}$ the (positive) eigenvalues of $-\Dsl^2$, the corresponding
$\ze$-function is defined and rewritten as a Mellin transform in the usual
way
\beq
\ze_\th(s):\equiv\ze_\th(-\Dsl^2,s):\equiv\sum_k \mu_k^{-s}=
{1\ov \Gamma(s)}\int\limits_0^\infty t^{s-1}\tr_\th(e^{-t(-\Dsl^2)})\ dt
\label{fdet.2}
\eeq
for ${\rm Re}(s)>d/2=1$ and its analytic continuation to ${\rm Re}(s)\leq 1$.


\subsection{Stepwise Calculation}

We will study how 
$\det_\th(i\Dsl_{\al,c})$ varies with $\th$, $\al$ and $c$ to
compute the normalized  determinant
\beq
{\det_\th(i\Dsl)\ov\det_0(i\pas)}\equiv
{\det_\th(i\Dsl_{\al=1,c})\ov\det_0(i\Dsl_{\al=0,0})}\ .
\label{fdet.30}
\eeq
The calculation is done in three steps. We shall calculate
all three factors in
\beq
{\det_\th(i\Dsl_{\al=1,c})\ov\det_0(i\Dsl_{\al=0,0})}\ \equiv\
{\det_\th(i\Dsl_{\al=1,c})\ov\det_0(i\Dsl_{\al=1,c})}\cdot
{\det_0(i\Dsl_{\al=1,c})\ov\det_0(i\Dsl_{\al=0,c})}\cdot
{\det_0(i\Dsl_{\al=0,c})\ov\det_0(i\Dsl_{\al=0,0})}
\label{fdet.3}
\eeq
in turn.
\par\noindent
From the generalized Feynman-Hellmann formula (\ref{hbbc.4}) and the fact that
$i\Dsl$ has no zeromodes so that the various partial
integrations are justified, the $\th$-variation of (\ref{fdet.2}) is found
to be
\beq
{d\ov d\th}\ze_\th(s)={2s\ov\Gamma(s)}\ \int\limits_0^\infty
t^{s-1}\ \tr_\th(e^{t\Dsl^2}\gaf)\ .
\label{fdet.5}
\eeq
Now one can use the asymptotic small-t-expansion for
$e^{t\Dsl^2}f$, where $f$ is a testfunction,
\beq
\tr_\th(e^{t\Dsl^2}f)={1\ov2\pi t}\sum_{m=0,1,\dots} t^{m/2}\
\tr_\th\left(\int a_{m/2}(f)+\oint b_{m/2}(f)\right)\ ,
\label{fdet.6}
\eeq
and where the $a_{m/2},b_{m/2}$ denote the corresponding volume and boundary
Seeley DeWitt coefficients respectively. Plugging this into the expression
(\ref{fdet.5}) yields \cite{O.22,O.23,O.24}
\beq
{d\ov d\th}{1\ov2}\log\det{}_\th(-\Dsl^2)=
-{1\ov4\pi}\int\tr(a_1(\gaf))-{1\ov4\pi}\oint\tr(b_1(\gaf))\ .
\label{fdet.7}
\eeq
For the squared Dirac operator $-\Dsl^2$ that part of $a_1$ which leads to
a nonvanishing $\gaf$-trace is known \cite{O.25} to be
$eF_{01}/2\pi$, i.e. independent of $\th$.
On the other hand $\oint b_1(.)$, which depends on the boundary
conditions, is calculated explicitly
in the appendix to be
\beq
\oint b_1(f)=\oint{1\ov2}\bigg\{
\left(\begin{array}{cc}
1&0\\0&1
\end{array}\right)
-{\log(e^\th)\ov\sinh(\th)}
\left(\begin{array}{cc}
e^\th&-1\\-1&e^{-\th}
\end{array}\right)
\bigg\}\ \pa_n f
\label{fdet.8}
\eeq
and does not contribute for $f=\gaf$.

\noindent
Integrating with respect to $\theta$ yields the 
following first factor in (\ref{fdet.3})
\beq
{\det_\th(i\Dsl_{\al=1,c})\ov\det_0(i\Dsl_{\al=1,c})}=
\exp\{-{\th\ov 2\pi}\int F_{01}\}=
\exp\{-{\th\ov2\pi}\int\lap\ph\}.
\label{fdet.10}
\eeq

To find the $\al$-variation leading to the
second factor we use (\ref{dadt.21}) 
in (\ref{fdet.2}) with the result
\beq
{d\ov d\al}\ze{}_\th(-\Dsl_{\al,c}^2,s)=
-{2s\ov\Gamma(s)}\int\limits_0^\infty
t^{s-1}\tr_\th(e^{t\Dsl_\al^2}2\ph\gaf),
\label{fdet.12}
\eeq
where we integrated by parts. Again we use the
small-$t$-expansion (\ref{fdet.6}) of the heat kernel, but now
with test function $f=2\ph\gaf$. Thus
\beq
{d\ov d\al}{1\ov2}\log\det{}_\th(-\Dsl_{\al,c}^2)=
{1\ov4\pi}\int\tr(a_1(2\ph\gaf))+{1\ov4\pi}\oint\tr(b_1(2\ph\gaf))
\label{fdet.13}
\eeq
where the universal $a_1(.)$ yields the wellknown Schwinger term \cite{O.25}.
Since now the normal derivative of the testfunction on the boundary
in non-zero, the last surface term contributes.
Using (\ref{fdet.8}) we end up with
\beq
{d\ov d\al}{1\ov2}\log\det{}_\th(-\Dsl^2_{\al,c})=
{1\ov4\pi}\int\limits_{M}2\ph\lap\ph
-{1\ov2\pi}\oint\limits_{\pa M}\log(e^\th)\ \pa_n\ph.
\eeq
Setting $\theta=0$ and integrating with respect to $\al$
yields the following second factor in (\ref{fdet.3}):
\beq
{\det_0(i\Dsl_{\al=1,c})\ov\det_0(i\Dsl_{\al=0,c})}=
e^{{1\ov2\pi}\int\ph\lap\ph}.
\label{fdet.14}
\eeq

We are left with the task to calculate the third factor
\beq
\log{\det_0(i\Dsl_{\al=0,c})\ov\det_0(i\Dsl_{\al=0,0})}=
-{1\ov2}\int\limits_0^c{d\ov ds}\bigg\vert_{s=0}{d\ov d\til c}\
\ze_0(-\Dsl_{\al=0,\til c}^2\;,s)\ d\til c\quad .
\label{fdet.16}
\eeq
For that we computed the heatkernel of the operator
\bdm
-\Dsl_{\al=0,\til c}^{\,2}=-\Big((\pa_0-2\pi i\til c/\be)^2+\pa_1^2
\Big)I_2
\edm
for $\th=0$. The explicit result is
\bea
K(t,x,y)&=&{1\ov4\pi t}\sum_{Z\times Z}(-1)^{m+n}
e^{-(\xi^0-n\be)^2/4t}e^{2\pi i\til c(\xi^0-n\be)/\be}
\nonumber
\\
{}&{}&\qquad\quad\left(\begin{array}{cc}
e^{-(\xi^1-2mL)^2/4t}&-e^{-(\eta-2m L)^2/4t}\\
-e^{-(\eta+2mL)^2/4t}&e^{-(\xi^1+2m L)^2/4t}
\end{array}\right)
\label{fdet.17}
\eea
which results in the trace ($V=\beta L$)
\beq
\int\limits_M \tr_{\!\th=0}\ (K(t,x,x))={V\ov 2\pi t}\Big(1+
\sum\nolimits\pri(-1)^{m+n}
e^{-{(n\be)^2+(2mL)^2\ov 4t}}\cos(2\pi n\til c)\Big)\ ,
\label{fdet.18}
\eeq
where the prime denotes the omission of the $(m,n)=(0,0)$ term.
The $\til c$-derivative of the Mellin transform, after substituting
$t\rightarrow1/t$, reads
\beq
{d\ov d\til c}\ze_0(-\Dsl^2_{\al=0,\til c}\;,s)=
{V\ov 2\pi\Gamma(s)}\int\limits_0^\infty\sum\nolimits\pri(-1)^{m+n}
e^{-t [(n\be/2)^2+(mL)^2]}\ {d\ov d\til c}\cos(2\pi n\til c)\ t^{-s}\ dt
\label{fdet.19}
\eeq
which may be integrated by parts (for $s>0$) to give
\bdm
{d\ov d\til c}\ze_0(-\Dsl^2_{\al=0,\til c}\;,s)=
-{2V\ov \pi}{s\ov\Gamma(s)}\int\limits_0^\infty\sum\nolimits\pri(-1)^{m+n}
{e^{-t\ [(n\be/2)^2+(mL)^2]}\ov [(n\be)^2+4(mL)^2]}
{d\ov d\til c}\cos(2\pi n\til c)\ t^{-s-1}\ dt\ .
\edm
Only the pole of order one of the integral
at $s=0$ can contribute to the $s$-derivative at $s=0$ of the
$\ze$-function. Since this pole entirely stems from the lower limit
of the integral we may split the latter into two parts
\bea
{d\ov d\til c}\ze_0 (-\Dsl^2_{\al=0,\til c}\;,s\downarrow 0)&=&
-{2V\ov \pi}{s\ov\Gamma(s)}
(\int_0^\ep\ldots+\int_\ep^\infty\ldots)
\nonumber
\\
&=&{2V\ov\pi}(s\!+\!\ga s^2\!+\!\ldots)\cdot\sum\nolimits\pri(-1)^{m+n}
{{d\ov d\til c}\cos(2\pi n\til c)\ov [(n\be)^2+4(mL)^2]} \ep^{-s}
+\ldots
\nonumber
\eea
to obtain
\beq
{d\ov ds}\bigg\vert_{s=0}{d\ov d\til c}\ze_0(-\Dsl_{\al=0,\til c}^2,s)=
{2V\ov\pi}\sum\nolimits\pri(-1)^{m+n}
{{d\ov d\til c}\cos(2\pi n\til c)\ov (n\be)^2+(2mL)^2}\ .
\label{fdet.20}
\eeq
Plugging this result into (\ref{fdet.16}) we end up with the
expression
\beq
-\Gamma(c)\equiv\log{\det_0(i\Dsl_{\al=0,c})\ov\det_0(i\Dsl_{\al=0,0})}
=-{V\ov\pi}\sum\nolimits\pri(-1)^{m+n}
{\cos(2\pi nc)-1\ov (n\be)^2+(2mL)^2}
\label{fdet.21}
\eeq
for the third factor of the functional determinant (\ref{fdet.30})
in the factorization (\ref{fdet.3}).
With the help of 
\bdm
{\hbox{Im}(\tau)\ov \pi}\;{\sum}^\pr\;{e^{2\pi i(ma_1+na_2)}\ov
\vert m+\tau n\vert^2}=-2\log\Big\vert{1\ov \eta(\tau)}\theta
\left[\begin{array}{c}\ha+a_1 \\ a_2\end{array}\right]\Big\vert
\edm
this result can be rewritten as \cite{Tata}
\beq
e^{-\Gamma(c)}={\det_0(i\Dsl_{\al=0,c})\ov\det_0(i\Dsl_{\al=0,0})}=
\left\{
\begin{array}{l}
{\theta_3(c,i\tau)\ov\theta_3(0,i\tau)}
\\
\\
e^{-\pi c^2/\tau}{\theta_3(ic/\tau,i/\tau)\ov\theta_3(0,i/\tau)}
\end{array}
\right .
\ .
\label{fdet.23}
\eeq
%
%
These two equivalent forms will be useful in the low- and high- 
temperature expansion of the condensates.


\subsection{Effective Action}

Now we can combine the classical (euclidean) action of the photon field,
rewritten in the new variables (\ref{dadt.11})
\beq
{1\ov4}F_{\mu\nu}F_{\mu\nu}={1\ov2e^2}\lap\ph\lap\ph\ \equiv\ :\
S_B[\ph]
\label{fdet.25}
\eeq
with our explicit result for the functional determinant (\ref{fdet.30}).
Collecting the contributions (\ref{fdet.10},\ref{fdet.14},\ref{fdet.23}) 
and adding the classical action (\ref{fdet.25}) we end
up with the effective action
\beq
\Gamma\equiv\Gamma_\th[c,\ph]\equiv N_f\Gamma(c)+\Gamma_\th[\ph]
\label{fdet.26}
\eeq
where $\Gamma(c)$ 
has been given in (\ref{fdet.23}) and $\Gamma_\th[\ph]$ is
\beq
\Gamma_\th[\ph]\equiv{1\ov2e^2}\bigg\{
\int\limits_M\ph\lap^2\ph-\mu^2\!\int\limits_M\ph\lap\ph+
\th\cdot \mu^2\int\limits_M\lap\ph\bigg\}
\label{fdet.27}
\eeq
and
\beq
\mu\;:\;\equiv\;\sqrt{N_fe^2\ov\pi}
\label{schwingermass}
\eeq
is the analog of the $\eta^\prime$-mass
in QCD.
We have used that the functional determinant
is the same for all flavours.
The functional measure takes the form
\beq
d\mu_\th[A]={1\ov Z_\th}\ e^{-\Gamma_\th[c,\ph]}
\ dc\ D\ph\ \de(\ch)\;D\ch\ .
\label{fdet.28}
\eeq
We have taken into account that the gauge-variation of
the Lorentz gauge-condition
\bdm
F:\equiv\pa_\mu A^\mu=\lap\ch
\edm
and the Jacobian of
the transformation from $\{A\}$ to the new variables $\{\ph,c,\ch\}$ are
independent of the fields. Actually, the corresponding
determinants cancel each other.

We conclude that the expectation value of any gauge-invariant operator $O$
(which will not depend on $\ch$) is given by
\beq
\big\langle O \big\rangle=
{\int dc\ D\ph\ \ O\ e^{-\Gamma_\th[c,\ph]}
\ov \int dc\ D\ph\ \ e^{-\Gamma_\th[c,\ph]}}
\label{fdet.29}
\eeq
with $\Gamma_\th[c,\ph]$ from (\ref{fdet.26},\ref{fdet.27}).


\section{Chiral Condensates}


Our result (\ref{fdet.29}) may be applied to calculate the chiral
condensates as
\bea
\<\ps\dag(x)P_\pm\ps(x)\>=
{\int dc\, D\ph\ \ S_\th(x,x)_{\pm\!\pm}\
e^{-\Gamma_\th[c,\ph]}
\ov
\int dc\, D\ph\ \
e^{-\Gamma_\th[c,\ph]}}
\label{rtpc.1}
\eea
with $S_\th$ from (\ref{fpro.11}) and $\Gamma_\th$ from (\ref{fdet.26}).
Both the (exponentiated) action and the Green's function factorize 
into parts which only
depend on $c$ and on $\ph$, respectively.
Thus (\ref{rtpc.1}) factorizes as
\beq
\<\ps\dag(x)P_\pm\ps(x)\>=C^{\pm}(x)\cdot D^{\pm}(x)
\label{rtpc.2}
\eeq
with $x^0$-independent factors
\bea
C^{\pm}(x^1)=
{\int dc\ \til S_\th(x,x)_{\pm\!\pm}\ e^{-N_f\Gamma(c)}
\ov\int dc\ \ e^{-N_f\Gamma(c)}}
\qquad,\qquad
D^{\pm}(x^1)=
{\int\ D\ph\ e^{\mp 2\ph(x)-\Gamma_\th[\ph]}
\ov\int\ D\ph\ \ e^{-\Gamma_\th[\ph]}}
\label{rtpc.4}
\eea
which depend on the parameters $\th, N_f, \be, L$.
Here and below the $c$-integrals extend over one period, e.g. $[-1/2,1/2]$.


\subsection{Harmonic Integral}

Now we shall see, how far we can evaluate the first factor in (\ref{rtpc.2})
which contains the integrals over the harmonic part of the gauge-field.

Plugging in the Green's function (\ref{fpro.6},\ref{fpro.11}) as well as
(\ref{fdet.21}) we obtain the unevaluated expression
\beq
C^{\pm}(x^1)=\pm{e^{\pm\th}\ov 4\pi L}
\sum_{Z\times Z}(-1)^{m+n}{\xi\!+\!m\ov(\xi\!+\!m)^2\!+\!(n\ta)^2}\cdot
{\int\limits_{-1/2}^{1/2}\cos(2\pi nc)
e^{-{N_f\ov2\pi}\sum\nolimits\pri(-1)^{k+l}
{\cos(2\pi l c)-1\ov k^2/\ta+l^2\cdot\ta} }\ dc\ov
\int\limits_{-1/2}^{1/2}
e^{-{N_f\ov2\pi}\sum\nolimits\pri(-1)^{k+l}
{\cos(2\pi l c)-1\ov k^2/\ta+l^2\cdot\ta} }\ dc}\ .
\label{hilf0}
\eeq
To investigate the low-temperature expansion we use
(\ref{fpro.8}) and the upper line in (\ref{fdet.23}) and arrive at
\beq
C^{\pm}(x^1)=\pm{e^{\pm\th}\ov4L}\sum_{n\in Z}
{(-1)^n\ov \sin(\pi[\xi-in\tau])}\cdot
{\int dc\
e^{\pm 2\pi i nc}\,
\;\theta_3^{N_f}(c,i\tau)\ov
\int dc\,\;\theta_3^{N_f}(c,i\tau)}\ ,
\label{hilf1}
\eeq
Alternatively, for the high-temperature
expansion we use (\ref{fpro.10}) and the lower line in (\ref{fdet.23}), so that
\beq
C^{\pm}(x^1)=\pm{e^{\pm\th}\ov2\be}\sum_{m\in Z}
{(-1)^m\ov\sinh(\pi[\xi+m]/\tau)}\cdot
{\int dc\ e^{-\pi c[N_fc\mp 2(\xi+m)]/\tau}\,
\;\theta_3^{N_f}(ic/\tau,i/\tau)\ov
\int dc\,e^{-\pi c^2N_f/\tau}\;\theta_3^{N_f}(ic/\tau,i/\tau)}\ .
\label{hilf2}
\eeq
For one flavour the $c$-integral in (\ref{hilf1}) is
easily calculated and one finds
\beq
C^{\pm}(x^1)=\pm {e^{\pm \theta}\ov 4L}\sum_{n\in Z}
{(-1)^ne^{-\pi\tau n^2}\ov \sin (\pi[\xi-in\tau])}\ .
\label{hilfs3}
\eeq


\subsection{Nonharmonic Integral}

Now we shall compute the second factor in (\ref{rtpc.2}) as defined in
(\ref{rtpc.4}). We recall that the integration extends over fields $\ph$,
which are periodic in the $x^0$ and satisfy Dirichlet boundary conditions at
the ends of the cylinder, i.e. at $x^1=0,L$.

Doing the gaussian integrals one ends up with
\beq
D^{\pm}(x^1)\;=\;\exp\Big\{{2\pi\ov N_f}K_{\mu^2}(x,x)\Big\}
\cdot
\exp\Big\{\pm{\th\ov2}\cdot\!\int\lap'K_{\mu^2}(x,x')
\pm{\th\ov2}\cdot\!\int\lap'K_{\mu^2}(x',x)\Big\}
\label{rtpc.7}
\eeq
where the integration is over $x'$ and the kernel
\beq
K_{\mu^2}(x,y)=
\<x\vert{\mu^2\ov-\lap(-\lap+\mu^2)}\vert y\>=
\<x\vert{1\ov-\lap}\vert y\>-\< x\vert{1\ov-\lap+\mu^2}\vert y\>
\label{rtpc.8}
\eeq
is with respect to Dirichlet boundary conditions. Being the
difference of two Green's functions with the same singular behaviour
it is finite at coinciding arguments.

The explicit form of the kernel is
\bea
K_{\mu^2}(x,y)\!&\!=\!&\!{V\ov \pi^2}\sum_{m,n\in Z}\nolimits^{\ '}
\bigg({1\ov(2 m L)^2\!+\!(n\be)^2}-
{1\ov(2 mL)^2\!+\!(n\be)^2\!+\!(\mu V/\pi)^2}\bigg)\cdot
\nonumber
\\
\nonumber
\\
&{}&\qquad\qquad\qquad
\cos\Big({2\pi m\xi^0\ov\be}\Big)\;
\sin\Big({\pi nx^1\ov L}\Big)\;\sin\Big({\pi ny^1\ov L}\Big)\ ,
\label{rtpc.9}
\eea
where the prime indicates the omission of the term with $m=n=0$.
For coinciding arguments $K$ becomes $x^0$-independent as required
by translational invariance.
For performing either the sum over $m$ or over $n$ in (\ref{rtpc.9})
one uses the formula
\bdm
\sum_{j\in Z}{\cos(jx)\ov j^2+a^2}=
{\pi\ov a}{\cosh(a(\pi-x))\ov\sinh(a\pi)}
\qquad\qquad(x\in[0,2\pi])
\edm
to end up either with the expression
\beq
K_{\mu^2}(x,x)={1\ov 2\pi}\sum\limits_{n\geq1}
\bigg({{\rm cth}(n\pi\tau)\ov n}-
(n\to \sqrt{n^2+(\mu L/\pi)^2})\bigg)
\Big(1-\cos(2\pi n \xi)\Big)\ ,
\label{rtpc.11}
\eeq
which is useful for the low temperature expansion, or alternatively with the
expression
\bea
K_{\mu^2}(x,x)&=&{1\ov 2\pi}
\sum\limits_{m\geq1}
{\cosh(m\pi/\tau)-\cosh(m\pi(1-2\xi)/\tau)
\ov m\ \sinh(m\pi/\tau)}
-(m\to \sqrt{m^2 +(\mu\be/2\pi)^2})
\nonumber
\\
\nonumber
\\
&{}&+\ {\xi(1-\xi)\ov2\tau}
+{\cosh(\mu L(1-2\xi))-\cosh(\mu L)
\ov 2\mu\be\ \sinh(\mu L)}\ ,
\label{rtpc.12}
\eea
which is useful for the high temperature expansion.
Both expressions (\ref{rtpc.11})
and (\ref{rtpc.12}) do indeed vanish as $x^1$ reaches the boundary in
accordance with the imposed boundary conditions.

Once we have the explicit formula (\ref{rtpc.9}) at hand we can 
compute in a straightforward way the expression
\beq
\int\lap_z K_{\mu^2}(z,x)\ dz=
-{4\ov\pi}\sum\limits_{n=1,3,\dots}
\big({1\ov n}-{n\ov n^2+(\mu L/\pi)^2}\big)\sin(\pi n \xi).
\label{rtpc.13}
\eeq
Applying the formula
\beq
\sum\limits_{n=1,3,\dots}{n\ \sin(nx)\ov n^2+a^2}
={\pi\ov4}{\rsh(a(\pi-x))+\rsh(ax)\ov\rsh(a\pi)}\qquad\quad(\ x\in\ ]0,\pi[\ )
\nonumber
\eeq
the expression (\ref{rtpc.13}) is seen to take the simple form
\beq
\int\lap_z K_{\mu^2}(z,x)\ dz=
{\sinh(\mu L (1-\xi))+\sinh(\mu L\xi)
\ov\sinh(\mu L)}-1\ .
\label{rtpc.14}
\eeq


\subsection{Final Result}

Now all pieces to compute the chiral condensate
(\ref{rtpc.2}) have been calculated.
For $C^\pm$ we have the two alternative
forms (\ref{hilf1}) and (\ref{hilf2}), and
$D^\pm$ is given by (\ref{rtpc.7})
wherein we can use one of the equivalent
representations (\ref{rtpc.11}) or
(\ref{rtpc.12}) for $K_{\mu^2}$ together
with (\ref{rtpc.14}). Thus we have
\bea
\<\psd P_\pm\ps\>(x^1)&=&
\pm{1\ov4L}\sum_{n\in Z}
{(-1)^n\ov \sin(\pi[\xi-in\tau])}\cdot
{\int dc\
e^{\pm 2\pi i nc}\,
\;\theta_3^{N_f}(c,i\tau)\ov
\int dc\,\;\theta_3^{N_f}(c,i\tau)}
\cdot
\nonumber
\\
&{}&
\exp\{{1\ov N_f}\sum\limits_{n\geq1}
\bigg({{\rm cth}(n\pi\tau)\ov n}-
(n\to \sqrt{n^2+({\mu L\ov\pi})^2})\bigg)
\Big(1-\cos(2\pi n \xi)\Big)\}
\cdot
\nonumber
\\
&{}&
\exp\{\pm\th\cdot{\sinh(\mu L (1-\xi))+\sinh(\mu L\xi)
\ov\sinh(\mu L)}\}
\label{rtpc.15}
\\
\<\psd P_\pm\ps\>(x^1)&=&
\pm{1\ov2\be}\sum_{m\in Z}
{(-1)^m\ov\sinh(\pi[\xi+m]/\tau)}\cdot
{\int dc\ e^{-\pi c[N_fc\mp 2(\xi+m)]/\tau}\,
\;\theta_3^{N_f}(ic/\tau,i/\tau)\ov
\int dc\,e^{-\pi c^2N_f/\tau}\;\theta_3^{N_f}(ic/\tau,i/\tau)}
\cdot
\nonumber
\\
&{}&
\exp\{{1\ov N_f}\sum\limits_{m\geq1}
{\cosh(m\pi/\tau)-\cosh(m\pi(1\!-\!2\xi)/\tau)
\ov m\ \sinh(m\pi/\tau)}
-(m\to \sqrt{m^2\!+\!({\mu\be\ov2\pi})^2})\}
\cdot
\nonumber
\\
&{}&
\exp\{{2\pi\ov N_f}\Big({\xi(1-\xi)\ov2\tau}
+{\cosh(\mu L(1-2\xi))-\cosh(\mu L)
\ov 2\mu\be\ \sinh(\mu L)}\Big)\}
\cdot
\nonumber
\\
&{}&
\exp\{\pm\th\cdot{\sinh(\mu L (1-\xi))+\sinh(\mu L\xi)
\ov\sinh(\mu L)}\}
\label{rtpc.16}
\eea
with excellent convergence properties in the low- and high-temperature
regime, respectively.

This result is one of the two
main results of this article. To simplify the
analysis we shall now study the condensates
at the midpoints of the cylinder.


\subsection{$\lan\psd P_{\pm}\ps\ran$ at Midpoints}

If a condensate survives at the midpoints when the boundaries 
are taken to infinity then the chiral symmetry is broken.
\newline
For $x^1=L/2$ the formulas (\ref{hilf1}), (\ref{hilf2}) simplify to
\bea
C^{\pm}({L\ov2})\!&\!=\!&\!
\pm{e^{\pm\th}\ov4 L}
\Bigg(1+2\sum\limits_{n\geq1}(-1)^n\;
{\int\cos(2\pi nc)\
\theta_3^{N_f}(c,i\tau) dc \ov
\cosh(n\pi\tau)
\int \theta_3^{N_f}(c,i\tau)\ dc }\ \Bigg)
\label{rtpc.19}
\\
C^{\pm}({L\ov2})\!&\!=\!&\!
\pm{e^{\pm\th}\ov\be}\!\sum\limits_{m\geq 0}(-1)^m
{\int\cosh((2m\!+\!1)\pi c/\tau)\,e^{-\pi N_f c^2/\tau}\theta_3^{N_f}(ic/\tau,
i/\tau)\,dc \ov
\sinh((2m\!+\!1)\pi/2\tau)\!
\int e^{-\pi N_f c^2/\tau}\theta_3^{N_f}(ic/\tau,i/\tau)\,dc }.
\label{rtpc.20}
\eea
The formulas (\ref{rtpc.11}) and (\ref{rtpc.12}) simplify to
\beq
K_{\mu^2}({L\ov2})={1\ov \pi}
\sum\limits_{n=1,3,\dots}
\bigg({{\rm cth}\big(n\pi\ta\big)\ov n}-
(n\to \sqrt{n^2+({\mu L\ov \pi})^2}\;)\bigg)
\label{rtpc.51}
\eeq
\beq
K_{\mu^2}({L\ov2})={1\ov 2\pi}
\sum\limits_{m\geq1}
\bigg({\rch\big(m\pi/\tau\big)-1\ov
m\;\rsh\big(m \pi/\tau\big)}-
(m\to \sqrt{m^2+({\mu \be\ov 2\pi})^2}\;)\bigg)
+{1\ov8\ta}-
{1\ov2 \mu\be}{\rch\big(\mu L\big)-1\ov
\rsh\big(\mu L\big)}.
\label{rtpc.52}
\eeq
Depending on which one of the equivalent forms (\ref{rtpc.11}) and
(\ref{rtpc.12}) for $K_{\mu^2}$ on the diagonal is used the factor
$D^{\pm}$ at the midpoints is found to read
\bea
D^{\pm}({L\ov2})\!&\!=\!&\!
\exp\Big\{{2\ov N_f}\sum\limits_{n=1,3,\dots}
{{\rm cth}(n\pi\tau)\ov n}
-(n\to \sqrt{n^2+({\mu L\ov \pi})^2}\ )\Big\}
\cdot
\nonumber
\\
{}\!&\!{}\!&\!
\exp\Big\{\mp\th\Big(1-{1/\rch({\mu L/2})}\Big)\Big\}
\label{rtpc.21}
\\
D^{\pm}({L\ov2})\!&\!=\!&\!
\exp\Big\{ {1\ov N_f}\sum\limits_{m\geq1}
{{\rm th}(m\pi /2\tau)\ov m}-
(m\to \sqrt{m^2+({\mu \be\ov 2\pi})^2}\ )\Big\}
\cdot
\nonumber
\\
{}\!&\!{}\!&\!
\exp\Big\{{\pi\ov N_f}\Big({1\ov4\ta}-
{{\rm th}(\mu L/2)\ov\mu\be}\Big)\Big\}
\cdot
\exp\Big\{\mp\th\Big(1-{1/\rch\big({\mu L/2}\big)}\Big)\Big\}
\label{rtpc.22}
\eea
which can be used to derive the low and high temperature expansions,
respectively.


\section{Noncommutativity of the Limits $\be^{-1}\rightarrow 0$ and
$ L\rightarrow\infty$}


In this section we show that the condensates at the midpoints,
$\<\ps\dag P_\pm\ps\>_\beta({ L\ov2})$, depend on the order in which the
limits $\beta\to\infty$ and $L\to \infty$ are taken; 
for $N_f=1$ the condensates survive only if we first let
$\beta\to\infty$.


\subsection{Limit $\be\rightarrow\infty$ for finite spatial length $ L$}

Here we derive the low temperature limit, i.e. the
condensates for $\be$
large compared to the fixed spatial length $L$ and the induced mass $\mu$.

From the explicit expression (\ref{rtpc.19}) we see at once that
\beq
C{^\pm}({L\ov2})=\pm{e^{\pm\th}\ov4L}(1+O(e^{-2\cdot\pi\be/2L}))
\label{nitl.4}
\eeq
for any number of flavours.

In order to get the corresponding limit for the second
factor $D^{\pm}({L\ov2})$ in (\ref{rtpc.2}) we use
(\ref{rtpc.21}) and perform the
asymptotic expansion of the coth to get
\bea
D^{\pm}({L\ov2})&=&
\exp\Big\{ {2\ov N_f}\sum\limits_{n=1,3,\dots}
\Big({1\ov n}-{1\ov\sqrt{n^2+(\mu L/\pi)^2}}\Big) \Big\}\cdot
\nonumber
\\
&{}&
\exp\Big\{ {4\ov N_f}\sum\limits_{n=1,3,\dots}
\sum\limits_{k\geq1}\Big({e^{-2k\cdot n{\pi\be\ov2L}}\ov n}-
{e^{-2k\sqrt{n^2+(\mu L/\pi)^2}{\pi\be\ov2L}}
\ov\sqrt{n^2+(\mu L/\pi)^2}}\Big) \Big\}\ \cdot
\nonumber
\\
&{}&
\exp\Big\{\mp\th\big(1-1/\rch(\mu L/2)\big)\Big\}
\label{nitl.5}
\eea
adapted to $\be\gg L$ as an intermediate result. Using the identity \cite{GrRy}
\beq
\sum\limits_{n=1,3\dots}{1\ov n}-{1\ov\sqrt{n^2+(x/\pi)^2\ }}=
{\ga\ov 2}+\ha\ln({x\ov\pi})-\sum\limits_{j\geq 1}(-)^j K_0(jx)
\label{nitl.6}\eeq
valid for $x>0$, where $\ga$ denotes the Euler Masceroni constant
and $K_0$ the zeroth Bessel function
the second factor can be rewritten as
\bea
D^{\pm}({L\ov2})&=&
e^{\ga/N_f}\Big(\mu L/\pi\Big)^{1/N_f}
\exp\Big\{ -{2\ov N_f}\sum\limits_{j\geq1}(-1)^j K_0(j\mu L) \Big\}
\ \cdot
\nonumber
\\
&{}&
\exp\Big\{\mp\th(1-1/\rch(\mu L/2))\Big\}\cdot
O(\exp({4\ov N_f}e^{-2\cdot\pi\beta/2L}))
\ .
\label{nitl.7}
\eea
Combining (\ref{nitl.4}) and (\ref{nitl.7}) we get the result
\bea
\<\ps\dag P_\pm\ps\>_{\beta}({ L\ov2})&=\ \;\pm\!\!&\!\!
{1\ov 4L}({\mu L\ov 2\pi})^{1/N_f}\,e^{\ga/N_f}
\exp\Big\{ -{2\ov N_f}\sum\limits_{j\geq1}
(-1)^j K_0(j\mu L)\; \Big\}\
\cdot
\nonumber
\\
\nonumber
\\
&{}&\!\!
\exp\Big\{ \pm\th/\rch(\mu L/2) \Big\}
\cdot\big(1+O\big(e^{-2\cdot\pi\be/2L}\big)\big)
\label{nitl.16}
\eea
where the $\th$ dependencies are found to cancel up to exponentially small
remainders.
In particular we have found a nonzero value for $\<\psd P_\pm\ps\>$ for
midpoints at zero temperature for any $N_f$ for finite spatial length $L$.


\subsection{Limit $ L\rightarrow\infty$ for finite temperature}

Here we give the large volume expansion of (\ref{rtpc.2}) valid for length 
$L$ which is large as compared to the fixed inverse temperature $\be$ and
$\mu^{-1}$.

The first task is to derive the high temperature asymptotics for
the first factor $C^{\pm}$ in (\ref{rtpc.2}).
By a variety of manipulations including infinite product representations
for the exponential factors constituting the measure we arrived at the
asymptotic result \cite{DuWiExpDual}
\beq
C^{\pm}({ L\ov2})=
\left\{
\begin{array}{lc}
\pm{e^{\pm\th}\ov\be}\cdot
{\sqrt{\be}\ov\pi\sqrt{2 L}}e^{-{5\ov2}{\pi L\ov\be}}
&(N_f=1)
\\
\\
\pm{e^{\pm\th}\ov\be}\cdot
2e^{-3{\pi L\ov\be}}
&(N_f=2)
\\
\\
\pm{e^{\pm\th}\ov\be}\cdot
4e^{-2{2N_f-1\ov N_f}{\pi L\ov\be}}
&(N_f\geq 3)
\end{array}
\right .
\label{nitl.13}
\eeq
which is an exponential decay which goes faster as the number of flavours
increases.

Also, we performed the asymptotic expansions of the hyperbolic
functions in (\ref{rtpc.22}) and arrived at
\bea
D^{\pm}({ L\ov2})&=&
\exp\Big\{ {1\ov N_f}\Big(\ga+{\pi\ov \mu\be}
+\ln({\mu\be\ov4\pi})
-2\sum\limits_{j\geq1}K_0(j\mu\be)\Big) \Big\}\cdot
\nonumber
\\
&{}&
\exp\Big\{ {2\ov N_f}\sum\limits_{m\geq1}\sum\limits_{ l\geq1}(-1)^ l
\Big({e^{- l\;m\pi /\tau}\ov m}-
{e^{- l \pi\sqrt{m^2+(\mu\be/2)^2}/\tau}\ov
\sqrt{m^2+(\mu\be/2)^2}}\Big) \Big\}\cdot
\nonumber
\\
&{}&
\exp\Big\{ {1\ov N_f}{\pi\ov4\tau}\Big(1-
{1+2\sum\limits_{ l\geq1}(-1)^ l
e^{- l \mu L}\ov
\mu L/2}\Big) \Big\}\cdot
\nonumber
\\
\nonumber
\\
&{}&
\exp\Big\{ \mp\th\pm2\th\sum\limits_{ l\geq0}(-1)^ l
e^{-(2 l+1)\mu L/2} \Big\}
\label{nitl.14}
\eea
where everything is at least exponentially suppressed as compared to the
growing factor in the second-last line.

Combining (\ref{nitl.13}) and (\ref{nitl.14}) we end up with the result
\bea
\<\ps\dag P_\pm\ps\>({L\ov2})&=\pm&
\left\{
\begin{array}{lc}
{1\ov\pi\sqrt{2\be L}}\cdot
e^{-{5\ov2}{\pi L\ov\be}}
&(N_f=1)
\\
{2\ov\be}\cdot
e^{-3{\pi L\ov\be}}
&(N_f=2)
\\
{4\ov\be}\cdot
e^{-2{2N_f-1\ov N_f}{\pi L\ov\be}}
&(N_f\geq 3)
\end{array}
\right\}
\cdot
e^{{1\ov N_f}{\pi L\ov2\be}}
\cdot
e^{\ga/N_f}
\cdot
\Big({\mu\be\ov4\pi}\Big)^{1/N_f}
\nonumber
\\
&{}&
\exp\Big\{ -{2\ov N_f}\sum\limits_{j\geq1}K_0(j \mu\be)
\Big\}
\cdot
O\Big(\exp\Big\{-{2\ov N_f}e^{-2\pi L/\be}\Big\}\Big)
\cdot
\\
&{}&
\exp\Big\{-{1\ov N_f}{2\pi\ov\mu\be}\sum\limits_{ l\geq1}(-1)^ l
e^{-l\mu L} \big\}
\cdot
\exp\Big\{ \pm2\th\sum\limits_{ l\geq0}(-1)^ l
e^{-(2 l+1)\mu L/2} \Big\}
\nonumber
\label{nitl.20}
\eea
which gives a decay
\beq
\<\ps\dag P_\pm\ps\>({ L\ov2})\sim
\left\{
\begin{array}{lc}
\pm{\rm const}\cdot{1\ov\sqrt{ L}}e^{-2{\pi L\ov\be}}&(N_f=1)
\\
\\
\pm{\rm const}\cdot e^{-{8N_f-5\ov2N_f}{\pi L\ov\be}}&(N_f\geq2)
\end{array}
\right .
\label{nitl.21}
\eeq
for spatial lengths $ L$ which are large compared to the inverse
temperature $\be$ and the inverse charge $e^{-1}$.


\subsection{Noncommutativity of the limits $\be\rightarrow\infty$ and
$ L\rightarrow\infty$}

Using the results of the previous subsections it is easy to show that
the two limits $\be\rightarrow\infty$ and $L\rightarrow
\infty$ do not commute.
\pan
Recall that $\<\psd P_\pm\ps\>({L\ov2})$ is a shorthand for
$\<\psd P_\pm\ps\>_{\th,N_f,\be,L}(x^1:={L\ov2})$.

Now the formulas (\ref{nitl.16}), (\ref{nitl.21}) imply
\bea
\lim_{ L\rightarrow\infty}\lim_{\be\rightarrow\infty}\
\<\ps\dag P_\pm\ps\>({ L\ov2})&=&
\left\{
\begin{array}{lc}
\pm{1\ov4\pi}e^\ga\sqrt{\,N_fe^2\ov\pi}&(N_f=1)
\\
\\
0&(\,N_f\geq2)
\end{array}
\right .
\label{nitl.23}
\\
\lim_{\be\rightarrow\infty}\lim_{ L\rightarrow\infty}\
\<\ps\dag P_\pm\ps\>({ L\ov2})&=&
0
\qquad\qquad\qquad\quad(\;\forall\;N_f\geq1)
\label{nitl.24}
\eea
respectively, which is the other main result of this paper.

From a physical point of view this means that the system under
consideration shows a distinctive hysteresis phenomenon: When both of $\be$
and $ L$ are sent to infinity, the one-flavour system keeps the knowledge of
which limit was performed first in the actual value of its chiral condensate.
Obviously there is no such non-commutativity for finite changes of the
lengths $\be$ and $ L$. We shall further comment
on this interesting behaviour in the conclusions.


\section{Discussion and Conclusions}


In this paper we have performed in a functional framework the quantization of
the $N_f$ flavour euclidean Schwinger model inside a finite temperature
cylinder with $SU(N_f)_A$ breaking local boundary conditions at the two spatial
ends to trigger chiral symmetry breaking.
We have determined the effective action for the bosonic subsystem subject to
these boundary conditions, which arises after integrating out the fermions.
We have shown the way the expectation value of an arbitrary gauge-invariant
operator can be computed and in particular we have performed the calculation
of the condensates  $\<\ps\dag P_\pm\ps\>(x)$ (to be
used as the most simple order parameters) for any point $x$ inside the
cylinder and any value of the inverse temperature $\be$ and spatial length
$ L$.

The quantization was greatly simplified by the fact that the boundary
conditions chosen (the CB-boundary-conditions) completely ban the zero modes.
Once more we emphasize the fact that our results have been obtained purely
analytically and without doing 'instanton physics'.
The technical aspects are rather different as those one encounters when
quantizing the theory on a sphere \cite{O.15} or on a torus
\cite{SaWi,O.14,O.33}.

Nevertheless our results are in full agreement with the earlier instanton-type
and small-quark-mass calculations.
Thus it seems that the CB-boundary-conditions applied at the two spatial ends
of the cylinder give a perfect substitute for introducing small quark masses
to trigger the chiral symmetry breaking and a real alternative to the study of
torons \cite{O.7} or fractons \cite{ShSm} or singular gauge-fields on $S^4$
\cite{O.8}.
The real advantage is of course the fact that they constitute almost exactly
the border of what can be calculated analytically. The functional integral over
the prepotential is gaussian, whereas, in general, the integration over the
harmonic part of the gauge-potential is not.
However the latter reduces to gaussian integrals in the low and high 
temperature expansions.

In the low temperature limit $\mu^{-1}=1/\sqrt{N_f e^2/\pi}\ll L\ll\be=T^{-1}$
we found for the chiral condensate the asymptotic value
\beq
\<\ps\dag P_\pm\ps\>({ L\ov2})=
\pm{1\ov4 L}e^{\ga/N_f}\bigg({\mu L\ov\pi}\bigg)^{1/N_f}=
\pm{1\ov4 L}e^{\ga/N_f}\bigg({\sqrt{N_fe^2/\pi\;}\; L\ov\pi}\bigg)^{1/N_f}
\label{conc.1}
\eeq
which, when restricted to the two-flavour case reduces to
\beq
\<\ps\dag P_\pm\ps\>({ L\ov2})=
\pm\bigg({e^\ga\sqrt{2e^2/\pi\;}\ov16\pi L}\bigg)^{1/2}\ .
\label{conc.2}
\eeq
This expression is identical to the result of Shifman and Smilga \cite{ShSm},
who allowed for fracton configurations on the torus.

In the high temperature limit $T=\be^{-1}\gg\sqrt{N_f e^2/\pi}\gg L^{-1}$ we
found for the chiral condensate an exponential decay with $T$.

For intermediate temperatures $T=\be^{-1}\simeq\sqrt{N_f e^2/\pi}$ and finite 
$ L$ one has to retreat to numerical methods to evaluate the remaining sum
and the integrals in (\ref{rtpc.19}) and (\ref{rtpc.21}) or equivalently in
(\ref{rtpc.20}) and (\ref{rtpc.22}).
One realizes that the observable $\<\ps P_\pm \ps\>$ viewed as a function
of $\,T$ strongly resembles the behaviour of an order parameter in a
system which suffers a second order phase transition for the case $N_f\geq2$.
However, the chiral condensate does not really vanish at any finite
temperature, it is just exponentially close to zero for temperatures larger
than the induced mass $\mu=\sqrt{N_f e^2/\pi}$.
Thus, in a strict sense, the chiral symmetry remains broken even for $N_f\geq2$
at all finite temperatures as long as $ L$ stays finite, as has been argued to
be a general fact by Dolan and Jackiw \cite{DoJa}. However, if $ L$ is sent
to infinity for finite $\be$, the condensate exponentially drops to zero.

Our main result is the fact that the limits
$\be\rightarrow\infty$ and $ L\rightarrow\infty$ do not commute for
the observable $\<\ps\dag P_\pm \ps\>$ in the $N_f\!=\!1$ case, since
\bea
\lim_{ L\rightarrow\infty}\lim_{\be\rightarrow\infty}\
\<\ps\dag P_\pm\ps\>({ L\ov2})&=&
\pm{1\ov4\pi}e^\ga\sqrt{e^2\ov\pi}
\qquad(\,N_f=1)
\label{conc.3}
\\
\lim_{\be\rightarrow\infty}\lim_{ L\rightarrow\infty}
\<\ps\dag P_\pm\ps\>({ L\ov2})&=&
0
\qquad\qquad\qquad\;(\;\forall\;N_f\geq1)
\label{conc.4}
\eea
which implies that there is no unique infinite volume limit.
Thus it seems that the combination of finite-temperature and CB- boundary
conditions provides an interesting tool for driving this system either into
the true or the wrong vacuum state.
The result (\ref{conc.3},\ref{conc.4}) is rather remarkable, since it means
that the one-flavour system shows some hysteresis phenomenon:
As far as we are aware of the literature, such phenomena are known for spin
systems but they are rather untypical for analytically solvable field theories.
However one of the interesting new results in this respect is the
work by Hetrick, Hosotani and Iso about the massive multi-flavour Schwinger
model on the zero temperature cylinder \cite{HHI}. They analyzed the situation
for small quark masses and finite (cyclic) spatial length $ L$. In particular
they found that the two limits $m\rightarrow 0$ and $ L\rightarrow\infty$
fail to commute. Thus we conclude that chirality breaking boundary conditions
give an interesting alternative to introducing small quark masses.


\section*{Acknowledgments}

One of the authors (S.D.) wishes to thank Daniel Wyler for his 
continued interest and many interesting discussions.
In addition useful conversations with Christian Wiesendanger and
Othmar Brodbeck are acknowledged.
\newline
This work has been supported by the Swiss National Science Foundation (SNF).


\section*{A $\;$ Explicit Construction of the Fermionic Heat Kernel}


In this appendix we sketch the construction of the heat kernel of the squared
Dirac operator $({i\Dsl\,\vert}_{\al=0})^2=(i\pas+2\pi c/\be \ga^0)^2$ on
a thermal manifold, which allows to compute the relevant Seeley DeWitt
coefficient, a task, which itself is postponed to appendix B. 
For that we construct the heat kernel $\til K$ 
on the finite cylinder $\{(x^0,x^1)\ \vert\ x^0\in [0,\be[\ ,
x^1\!\geq 0\  \}$ which obeys (besides the usual heat kernel relations)
the boundary conditions
\bea
(B_\th \til K)(t,x^0,0,y)&=&\til K(t,x^0,0,y)\label{A.1}\\
(B_\th i\pas \til K)(t,x^0,0,y)&=&(i\pas \til K)(t,x^0,0,y)\label{A.2}\\
\til K(t,x^0\!+\!\be,x^1,y)&=&-\til K(t,x^0,x^1,y)\label{A.3}
\eea
as well as the adjoint relations with respect to $y$, where $B_\th$ is a
shorthand for $B_L(\th)$ defined in (\ref{hbbc.6}).
\pan
The trick is to start considerations on the half plane
$\{(x^0,x^1)\ \vert\ x^1\geq 0\}$, since here the above squared Dirac
operator can be decomposed as
\beq
(i\pas_x +{2\pi c/\be}\cdot\si^1)^2=e^{2\pi ic x^0/\be}
(i\pas_x)^2 e^{-2\pi ic x^0/\be}
\label{A.4}
\eeq
and correspondingly the free heat kernel takes the simple form
\beq
{1\ov 4\pi t}e^{-((\xi^0)^2+(\xi^1)^2)/4t}
e^{2\pi ic\xi^0/\be}=
{1\ov 4\pi t}e^{-((\xi^0-4\pi ict/\be)^2+(\xi^1)^2)/4t}
e^{-4\pi^2c^2t/\be^2}
\label{A.5}
\eeq
where $\xi^0=x^0-y^0, \xi^1=x^1-y^1$.
Using that the kernel can be Fourier transformed and from the
mirror principle one is led to consider the expression
\bdm
{1\ov(2\pi)^2}\int\limits_{-\infty}^{\infty}\!\int\limits_{-\infty}^{\infty}
e^{-(k_0^2+k_1^2)t} e^{ik_0\xi^0+ik_1\xi^1}\ dk_0dk_1
\edm
\bdm
+{1\ov(2\pi)^2}
\int\limits_{-\infty}^{\infty}\!\int\limits_{-\infty}^{\infty}
e^{-(k_0^2+k_1^2)t}
\left(\begin{array}{cc}
f(k_0,k_1)&g(k_0,k_1)\\
g(k_0,k_1)&h(k_0,k_1)
\end{array}\right)
e^{ik_0\xi^0+ik_1\eta} \ dk_0dk_1
\edm
as an ansatz for the heat kernel of the operator $(i\pas)^2=-\Delta\cdot I_2$
on the half plane. The boundary condition at $x^1=0$ immediately transforms
into an algebraic relation among $f,g,h$ which is solved by the expressions
\bea
f(k_0,k_1)&=&
-{e^{2\th}(k_0\!-\!ik_1)-(k_0\!-\!ik_1)\ov
e^{2\th}(k_0\!+\!ik_1)-(k_0\!-\!ik_1)}\nonumber\\
g(k_0,k_1)&=&
-{2e^\th\;ik_1\ov
e^{2\th}(k_0\!+\!ik_1)-(k_0\!-\!ik_1)}\nonumber\\
h(k_0,k_1)&=&
-{e^{2\th}(k_0\!+\!ik_1)-(k_0\!+\!ik_1)\ov
e^{2\th}(k_0\!+\!ik_1)-(k_0\!-\!ik_1)}\nonumber\quad.
\eea
The resulting integrals can be done in two steps. First only the numerators
of the functions $f,g,h$ are taken into account and the resulting expressions
are integrated over. Second the full expressions have to be read as differential
equations in $x^0,x^1$ in the manner indicated by the previously omitted
denominators of the functions $f,g,h$. There is a unique solution to this
procedure which falls off in both $x^0$ plus the positive $x^1$ directions
(note $\th\in\mbox{R}$) :
\bdm
\begin{array}{l}
{1\ov 4\pi t}\uad e^{-{(\xi^0)^2+(\xi^1)^2\ov 4t}}+{1\ov 4\pi t}
\left(\begin{array}{cc}e^\th\rsh \th&-\rch \th\\
-\rch \th&-e^{-\th}\rsh \th\end{array}\right)
e^{-{(\xi^0)^2+\eta^2\ov 4t}}
\\
\\
+{i\ov 8\pi^{1/2}t^{3/2}}
\left(\begin{array}{cc}e^\th\rsh \th&-\rsh \th\\
-\rsh \th&e^{-\th}\rsh \th\end{array}\right)
\cdot\left( \xi^0\rch\th +i\eta\;\rsh\th \right)
\uad\cdot
\\
\\
\quad\!\!\,
e^{-{(\xi^0\mbox{\small ch$\th$}+i\eta\;\mbox{\small sh$\th$})^2
\ov 4t}}\cdot\left( 1+\mbox{erf}
({i\xi^0\mbox{sh$\th$}-\eta\;\mbox{ch$\th$}\ov 2t^{1/2}})
\right)\quad.
\end{array}
\edm
Since on the half-plane the operator $(i\pas+2\pi c\si^1/\be)^2$ has the
decomposition (\ref{A.4}) this immediately yields its heat kernel ( to
be denoted $\til K$ ) by just including a factor $e^{2\pi ic\xi^0/\be}$
in each term. Finally the finite temperature boundary condition (\ref{A.3})
is taken into account by substituting $\xi^0$ by
$\xi^0\!-\!n\be$, including an additional 
$(-1)^n$ and performing the sum over $n\in\mbox{Z}$.

The heat kernel ${\til K}_{\al=0}$ of $(i\Dsl\,\vert_{\al=0})^2=(i\pas+2\pi
c\si^1/\be)^2$ subject to the boundary conditions (\ref{A.1}) - (\ref{A.3})
on the half cylinder $\{\ (x^0,x^1)\ \vert\ x^0\!\in [0,\be[\ ,x^1\geq 0\ \}$
takes the final form
\bea
{\til K}&=&\sum (-1)^n
{1\ov 4\pi t}\uad e^{-{(\xi^0\!-n\be)^2+(\xi^1)^2\ov 4t}}\uad
e^{2\pi ic(\xi^0\!-n\be)/\be}
\nonumber
\\
\nonumber
\\
&+&\sum (-1)^n
{1\ov 4\pi t}\uad
\left(\begin{array}{cc}e^\th\rsh \th&-\rch \th\\
-\rch \th&-e^{-\th}\rsh \th\end{array}\right)
e^{-{(\xi^0\!-n\be)^2+\eta^2\ov 4t}}\uad
e^{2\pi ic(\xi^0\!-n\be)/\be}
\nonumber
\\
\nonumber
\\
&+&\sum (-1)^n
{i\ov 8\pi^{1/2}t^{3/2}}\uad
\left(\begin{array}{cc}e^\th\rsh \th&-\rsh \th\\
-\rsh \th&e^{-\th}\rsh \th\end{array}\right)
\cdot\left( (\xi^0\!\!-\!n\be)\rch \th
+i\eta\;\rsh \th \right)\cdot
\nonumber
\\
& &e^{-{\left((\xi^0\!-n\be)\mbox{\small ch$\th$}+
i\eta\;\mbox{\small sh$\th$}\right)^2
\ov 4t}}\cdot\left( 1+\mbox{erf}
({i(\xi^0\!\!-\!n\be)\mbox{sh$\th$}\!-\!\eta\;\mbox{ch$\th$}
\ov 2t^{1/2}})\right)
\label{A.6}
\eea
where the sums run over $n\!\!\in$Z and can be seen to converge absolutely and
thus uniformely.


\section*{B $\;$ Extraction of the Relevant Heat Kernel Coefficients}


In this appendix we shall compute the surface Seeley DeWitt coefficient $b_1$
of the operator $-\Dsl^2$ which enters the calculation of it's functional
determinant. We first note that in general $\oint\mbox{tr}\left(b_m(\vp)
\right)$ with a smooth test function $\vp$ on a $d$ dimensional manifold $M$
has the expansion
\bdm
\oint\limits_{\pa M}\mbox{tr}\left(b_{m}(\vp)\right)=
\sum\limits_{p=0}^{d-1}\oint\limits_{\pa M}\mbox{tr}
\left( b_{m.p}(R,\ch,F_{..})\cdot\pa_n^p\vp\right)\ ,
\edm
where $b_{m.p}$ is a gauge-invariant and Lorentz-covariant local polynomial
in the intrinsic and extrinsic curvatures of the boundary as well as in the
field strength and its covariant derivatives on the boundary. Here
$\pa_n^p\vp$ denotes the $p$ fold derivative of the test function $\vp$
along the (outward oriented) normal of the boundary. In the case of a two
dimensional manifold with Hrasko Balog boundary conditions (\ref{hbbc.6}) the
expansion of $\oint\mbox{tr}\left( b_1(\vp)\right)$ simplifies to
\bdm
\oint\mbox{tr}\left(b_1(\vp)\right)=
\oint\mbox{tr}\left(b_{1.0}(\th)\,\ch\cdot\vp\right)+
\oint\mbox{tr}\left(b_{1.1}(\th)\cdot\pa_n\vp\right)\ .
\edm
For our purposes it is sufficient to know the coefficient $b_{1.1}$, since the
first term does not contribute to (\ref{fdet.13}) (due to $\vp :\equiv H+H\dag
\,= 0$ on $\pa M$) and in (\ref{fdet.7}) it would yield an uninteresting
constant which finally cancels in expectation values of gauge-invariant
operators.
The function $b_{1.1}$ can be determined from the heat kernel on the diagonal,
$K(t,x,x)$ of $-\Dsl^2=-\Dsl^2\vert_{\al=1}$ which is identical
to $\til K(t,x,x)$ of $-\Dsl^2\vert_{\al=0}$
by calculating
\bdm
\int\limits_M K(t,x,x)\cdot \vp(x)=\int\limits_M \til K(t,x,x)\cdot \vp(x)\sim
\int\limits_0^\infty \til K(t,x,x)\cdot
\left(\vp(x^0,0)+x^1\cdot\pa_1\vp(x^0,0)+...\right)dx^1
\edm
where $\til K$ denotes the heat kernel (\ref{A.6}) calculated in appendix A.
In writing this expansion we have anticipated that for small $t$ the 
heat kernel on the diagonal is sharply peaked about the boundary whereupon it
is justified to expand the test function $\vp$ about $x^1=0$. 
Using this result and denoting $\vp'(x^0,.)$ the first derivative of $\vp$ with
respect to it's second argument one has to compute an expression whose
first few terms in the small $t$ expansion take the form
\bea
\sum\limits_{n\in Z}(-1)^n
\left(\begin{array}{cc}1&0\\0&1\end{array}\right)
&\cdot&{1\ov 4\pi t}\ e^{-{n^2\be^2\ov 4t}}\,e^{-2\pi inc}\cdot
\int\limits_0^\infty\,\,\vp(x^0,x)\,dx\nonumber\\
+\sum\limits_{n\in Z}(-1)^n
\left(\begin{array}{cc}e^\th\rsh &-\rch \\
-\rch &-e^{-\th}\rsh \end{array}\right)
&\cdot&{1\ov 4\pi t}\,e^{-{n^2\be^2\ov 4t}}\,e^{-2\pi inc}\cdot
\int\limits_0^\infty e^{-{x^2/t}}dx\cdot \vp(x^0,0)\nonumber\\
+\sum\limits_{n\in Z}(-1)^n
\left(\begin{array}{cc}e^\th\rsh &-\rch \\
-\rch &-e^{-\th}\rsh \end{array}\right)
&\cdot&{1\ov 4\pi t}\,e^{-{n^2\be^2\ov 4t}}\,e^{-2\pi inc}\cdot
\int\limits_0^\infty x\ e^{-{x^2/t}}dx\cdot \vp'(x^0,0)\nonumber\\
+\sum\limits_{n\in Z}(-1)^n
\left(\begin{array}{cc}e^\th\rsh &-\rsh \\
-\rsh &+e^{-\th}\rsh \end{array}\right)
&\cdot&\int\limits_0^\infty
1\,{\rch n\be-2i\rsh x\ov 8\pi^{1/2} i t^{3/2}}\,
e^{-{(\mbox{\small ch$n\be$}-\mbox{\small$2i$sh$x$})^2\ov 4t}}\uad\cdot
\nonumber\\
& &\quad
\left(1-\mbox{erf}({\rch 2x+i\rsh n\be\ov 2t^{1/2}})\right)\,
e^{-2\pi inc}\, dx\cdot \vp(x^0,0)
\nonumber\\
+\sum\limits_{n\in Z}(-1)^n
\left(\begin{array}{cc}e^\th\rsh &-\rsh \\
-\rsh &+e^{-\th}\rsh \end{array}\right)
&\cdot&\int\limits_0^\infty
x\,{\rch n\be-2i\rsh x\ov 8\pi^{1/2} i t^{3/2}}\,
e^{-{(\mbox{\small ch$n\be$}-\mbox{\small$2i$sh$x$})^2\ov 4t}}\uad\cdot
\nonumber\\
& &\quad
\left(1-\mbox{erf}({\rch 2x+i\rsh n\be\ov 2t^{1/2}})\right)\,
e^{-2\pi inc}\, dx\cdot \vp'(x^0,0)\nonumber
\eea
where the first line gives the usual $a_0$ coefficient whereas the remaining
four integrals contain information about the $b_{1/2}$ and $b_1$ coefficients.
Here and below we use the abbreviations $\rsh=\rsh\th,\rch=\rch\th$.

The first and second integrals are easily evaluated using the formulas
\bea
I_1:=\int\limits_0^\infty e^{-{x^2\ov t}}\,dx={\sqrt{\pi t}\ov 2}\quad,
\qquad
I_2:=\int\limits_0^\infty x\cdot e^{-{x^2\ov t}}\,dx={t\ov 2}\quad.
\nonumber
\eea
The third and fourth integrals are handled using the formulas
\bea
I_3:&=&\int\limits_0^\infty
{\rch n\be-2i\rsh x\ov 8\pi^{1/2}it^{3/2}}\
e^{-{(\mbox{\small ch$n\be$}-\mbox{\small $2i$sh$x$})^2\ov 4t}}\,
(1-\mbox{erf}({\rch 2x+i \rsh n\be\ov 2t^{1/2}}))\, dx
\nonumber\\
&=&-{1\ov 8\pi^{1/2}t^{1/2}}\,{\rch \ov\rsh }\,e^{-{n^2\be^2\ov 4t}}
+{1\ov 8\pi^{1/2}t^{1/2}}\,{1\ov\rsh }\cdot
e^{-{{\rm ch}^2\ n^2\be^2\ov 4t}}
\mbox{erfc}({i\rsh n\be\ov 2t^{1/2}})
\nonumber\\
\nonumber\\
I_4:&=&\int\limits_0^\infty
\, x\ {\rch n\be-2i\rsh x\ov 8\pi^{1/2}it^{3/2}}\
e^{-{(\mbox{\small ch$n\be$}-\mbox{\small $2i$sh$x$})^2\ov 4t}}\,
(1-\mbox{erf}({\rch 2x+i\rsh n\be\ov 2t^{1/2}}))\, dx
\nonumber\\
&=&-{1\ov 8\pi}\, {\rch \ov\rsh }\, e^{-{n^2\be^2\ov 4t}}
+{1\ov 8\pi^{1/2}t^{1/2}}\, {1\ov\rsh }\cdot \int\limits_0^\infty
e^{-{(\mbox{\small ch$n\be$}-\mbox{\small $2i$sh$x$})^2\ov 4t}}
\mbox{erfc}({\mbox{\small ch}2x+i\mbox{\small sh}n\be\ov 2t^{1/2}})\, dx
\nonumber
\eea
which result in the small $t$ asymptotics
\beq
I_3\;\cong\;
\left\lbrace
\begin{array}{ll}
\vspace{0.3cm}
-{1\ov 8\pi^{1/2}t^{1/2}}{\rch\ov\rsh}\ e^{-{n^2\be^2\ov 4t}}
-{i\ov 4\pi\rsh^2} e^{-{n^2\be^2\ov 4t}}(1+O(t)) 
& \hspace{0.44cm} (n>0) \\
\vspace{0.2cm}
-{1\ov 8\pi^{1/2}t^{1/2}}{\rch\ov\rsh}+{1\ov 8\pi^{1/2}t^{1/2}}{1\ov\rsh}
& \hspace{0.44cm} (n=0) \\
-{1\ov 8\pi^{1/2}t^{1/2}}{\rch\ov\rsh}\ e^{-{n^2\be^2\ov 4t}}
+{i\ov 4\pi\rsh^2} e^{-{n^2\be^2\ov 4t}}(1+O(t)) 
& \hspace{0.44cm} (n<0)
\end{array}
\right .
\label{B.1}
\eeq
\vspace{0.3cm}
\beq
I_4\;\cong\;
\left\lbrace
\begin{array}{ll}
\vspace{0.3cm}
-{1\ov 8\pi}{\rch\ov\rsh}\ e^{-{n^2\be^2\ov4t}}
-{it^{1/2}\ov 8\pi^{1/2}\rsh^2 n\be}\ e^{-{n^2\be^2\ov4t}}(1+O(t^{1/2}))
& (n>0)\\
\vspace{0.2cm}
-{1\ov 8\pi}{\rch\ov\rsh}+{1\ov 8\pi}
{\log(\rch+\rsh)-\log(\rch-\rsh)\ov 2\rsh^2} 
& (n=0)\\
-{1\ov 8\pi}{\rch\ov\rsh}\ e^{-{n^2\be^2\ov4t}}
-{it^{1/2}\ov 8\pi^{1/2}\rsh^2 n\be}\ e^{-{n^2\be^2\ov4t}}(1+O(t^{1/2}))
& (n<0)
\end{array}
\right .
\label{B.2}
\eeq
where the result for $I_3$ immediately follows from the asymptotic expansion
\cite{AbSt}
\beq
\sqrt{\pi}\ z\ e^{z^2}\mbox{erfc}(z)\cong
1+\sum_{k=1}^\infty (-1)^k {1\cdot 3\cdot\,\ldots\,\cdot(2k\!-\!1)
\ov (2^k z^{2k})} \qquad
(z\rightarrow\infty,\vert\mbox{arg} z\vert < {3\pi\ov 4})
\label{B.3}
\eeq
whereas the expression for $I_4$ results from a computation establishing
the asymptotic behaviour
\bea
f(w)&=&\int\limits_0^\infty
e^{-(\rch w-i\rsh x)^2}
\mbox{erfc}(\rch  x+i\rsh  w) dx
\nonumber
\\
&=&e^{-w^2}\Big(
-{i\ov 2\rsh}\cdot{1\ov w}
+{\rch\ov2\pi^{1/2}\rsh^2}\cdot{1\ov w^2}
+{i\ov 4\rsh}\cdot{1\ov w^3}
+O({1\ov w^4})
\Big)
\label{B.4}
\eea
for $w\gg 1$.
\pan
Putting everything together we arrive at the small $t$ expansion of the
heat kernel
\beq
\int\limits_M K(t,x,x)\vp(x)\,dx-O(t^{1/2})=
\label{B.5}
\eeq
\bdm
\begin{array}{l}
+{1\ov4\pi t}\cdot\bigg\{ \sum\limits_{\bf Z}(-1)^n
e^{-{n^2\be^2\ov4t}}\cos(2\pi nc) \bigg\}\cdot\int\vp(x^0,x^1)\;d^2x
\\
\\
+{1\ov8\pi^{1/2}t^{1/2}}\cdot\bigg\{
\bigg(\begin{array}{cc} -1&0\\0&-1 \end{array}\bigg)
\sum\limits_{\bf Z}(-1)^n e^{-{n^2\be^2\ov4t}}\cos(2\pi nc)
+\bigg(\begin{array}{cc} e^\th&-1\\-1&e^{-\th} \end{array}\bigg)
\bigg\}\cdot\int\vp(x^0,0)\;dx^0
\\
\\
+{1\ov2\pi}\cdot\bigg\{ {1\ov\rsh(\th)}
\bigg(\begin{array}{cc} e^\th&-1\\-1&e^{-\th} \end{array}\bigg)
\sum\limits_{n\geq1}(-1)^n e^{-{n^2\be^2\ov4t}}\sin(2\pi nc)
\bigg\}\cdot\int\vp(x^0,0)\;dx^0
\\
\\
+{1\ov8\pi}\cdot\bigg\{
\bigg(\begin{array}{cc} -1&0\\0&-1 \end{array}\bigg)
\sum\limits_{\bf Z}(-1)^n e^{-{n^2\be^2\ov4t}}\cos(2\pi nc)
+{\ln(e^\th)\ov\rsh(\th)}
\bigg(\begin{array}{cc} e^\th&-1\\-1&e^{-\th} \end{array}\bigg)
\bigg\}\cdot\int\vp'(x^0,0)\;dx^0
\end{array}
\edm
which is used to determine the effective action (\ref{fdet.26}).
This formula resolves also the apparent paradox that
the $\th$-term in the effective action (\ref{fdet.26}) is linear,
whereas the whole model was defined through hyperbolic functions
of $\th$, thus there must be an invariance under the replacement
$\th\rightarrow\th+2\pi i$.


\clearpage

\end{document}